\documentclass[prc]{revtex4}%
\usepackage{amsfonts}
\usepackage{amsmath}
\usepackage{amssymb}
\usepackage{graphicx}%
\providecommand{\U}[1]{\protect\rule{.1in}{.1in}}
\newtheorem{theorem}{Theorem}
\newtheorem{acknowledgement}[theorem]{Acknowledgement}

\begin{document}
\title{Description\ of\ Superdeformed\ Nuclei\ in\ the\ $A\sim190$%
\ Region\ by\ Generalized\ Deformed$\ su_{\mathbf{q}}\left(  2\right)  $}
\author{H. H. Alharbi}
\email{alharbi@kacst.edu.sa}
\affiliation{National Center for Mathematics and Physics, KACST, P.O. Box 6086, Riyadh
11442, Saudi Arabia, }
\author{H. A. Alhendi}
\email{alhandi@ksu.edu.sa / hhalhendi@hotmail.com}
\author{F. S. Alhakami}
\affiliation{Department of Physics and Astronomy, College of Science, King Saud University,
P.O. Box 2455, Riyadh 11454, Saudi Arabia.}

\begin{abstract}
The generalised deformed $su_{\mathbf{q}}\left(  2\right)  \ $model is applied
to 79 superdeformed bands in the region $A\sim190$. The transition energies
and the moments of inertia are calculated within the model, Its validity is
investigated by comparing it with the experimental data. Both the standard
$su_{\mathbf{q}}\left(  2\right)  $ and the generalized one fail to account
for the uprising and the downturn of the dynamic moment of inertia. Both
models, however, show remarkable agreement with the available experimental
data at low angular frequancy $(\hbar\omega\leqslant0.25MeV).$

\end{abstract}
\volumeyear{year}
\volumenumber{number}
\issuenumber{number}
\startpage{101}
\endpage{102}
\maketitle

\section{INTRODUCTION}

\noindent Superdeformed nuclei were first observed in fission isomers in the
actinide region \cite{Polikanov}. A theoretical explanation of the occurrence
of fission isomers, based on shell effect corrections on the liquid drop
potential energy surface, was, at then, offered by Strutinsky
\cite{Strutinsky}. The main result was the possible existence of a second
minimum in the potential energy as function of nuclear deformation. It is
expected nowadays that a third minimum may occur corresponding to
hyper-deformed nuclei \cite{Galindo}.

\noindent A superdeformed rotational band in $^{152}$Dy in the form of series
of $\gamma$-ray energies was first populated in the heavy-ion
fusion-evaporation reaction $^{108}$Pd( $^{48}$Ca , 4n ) $^{152}$Dy
\cite{Twin}. Since then extensive experimental and theoretical studies have
been undertaken. At present superdeformed bands have been observed in various
atomic mass region \cite{Singh}. Most notable regions are at $A\sim130$, $150
$, and $190$ in which axes ratios are,respectively, close to $1.5:1$, $2:1$,
and $1.7:1$ \cite{Ward}.

\noindent Superdeformed nuclei enjoy several characteristics that make them of
particular interest theoretically and experimentally. For beside their extreme
shape and stability against fission they show great regularity in their
rotational bands and exhibit some type of universal phenomenon in relation to
the existence of nearly identical bands in pairs of nuclei in different mass
region and as a result their dynamic moments of inertia are approximately
similar \cite{Baktash}. It is expected that the process of the decay of
superdeformed nuclei to normal deformed nuclei could proceed through quantum
tunneling \cite{Stafferd}.

\noindent For high-spin, superdeformed rotational spectra follow, in general,
approximately that of a rigid rotor. Hence the kinematics and the dynamic
moments of inertia are nearly constant with slight gradual increase with
angular momentum, at low angular frequency. At high angular frequency the
dynamic moment of inertia shows irregular behavior.

\noindent In this work we consider the q-deformation of the enveloping Lie
algebra $su_{q}\left(  2\right)  $ \cite{Biedenharn}, which has recently
attracted much interest for the calculation of rotational spectra of deformed
\cite{Raychev} and superdeformed nuclei \cite{Bonatsos}. The validity of the
standard $su_{q}\left(  2\right)  $ model has, however, been recently
questioned \cite{Meng}. A generalized form of the model which is obtained by
replacing the angular momentum spectral expression $I(I+1)$ by $I(I+c)$ has
been used to describe successfully the vibrational, transitional and the
rotational nuclear spectra of well deformed nuclei \cite{Bonatsos2}. Here we
apply this generalized form to the calculation of the rotational transition
energies , the kinematic moments of inertia and the dynamic moments of inertia
for 79 superdeformed energy bands in the region $A\sim190 $, and compare it
with the experimental data. A sensitive measure of the applicability of a
model to superdeformed bands is the dynamic moment of inertia. This is becuse,
it is inersely proportional to the difference of the transition energies and
these transition energies are closly spaced. The model shows remarkable
agreement with the experimental data in the rotational region at low angular
frequency ($\hbar\omega\leqslant0.25$ $MeV$). A comparison with the standard
$su_{q}\left(  2\right)  $ model is also made. It is also shown that in
addition to the previously predicted deviation of the standard $su_{q}\left(
2\right)  $ in the case of deformed nuclei, it does so for the case of the
superdeformed nuclei considered in this work. It is also concluded, contrary
to the expectation of reference \cite{Bonatsos2}, that in the rotational
region the \noindent generalized $su_{q}\left(  2\right)  $ does not in
general coincide with the standard one.

\noindent In the following section we present a brief description of the model
and in the next section we present our results and conclusion.

\section{\bigskip MODEL DESCRIPTION}

The $su_{q}\left(  2\right)  \ $algebra is a $q$ deformation of the $SU\left(
2\right)  $ Lie algebra and is generated by the operators, $J_{-}$, $J_{0}$,
and $J_{+}$, which obey the commutation relations \cite{Biedenharn}%
\cite{Raychev}:%

\begin{equation}
\left[  J_{0},J_{\pm}\right]  =\pm J_{\pm},\mathrm{\;\;\;\;}\left[
J_{+},J_{-}\right]  =\left[  2J_{0}\right]  ,\label{comut}%
\end{equation}

with $J_{0}^{\dagger}$ = $J_{0}$, $J_{+}^{\dagger}$= $J_{-}$ and $\left[
x\right]  $ is the $q$ number defined as%

\begin{equation}
\left[  x\right]  =\frac{q^{x}-q^{-x}}{q-q^{-1}}\label{GrindEQ2}%
\end{equation}

In terms of the parameterization $\tau=\ln q$, this equation takes the form:%

\begin{equation}
\left[  x\right]  =\frac{e^{\tau x}-e^{-\tau x}}{e^{\tau}-e^{-\tau}}%
=\frac{\sinh\tau x}{\sinh\tau}\label{GrindEQ3}%
\end{equation}
In the $su_{q}\left(  2\right)  $ formalism it is suggested that rotational
spectra of nuclei can be well described by a Hamiltonian proportional to the
second-order Casimir operator of the quantum algebra of $su_{q}\left(
2\right)  $ in a manner similar to that of the $SU\left(  2\right)  $ rotator algebra..

The second-order Casimir operator of $su_{q}\left(  2\right)  $ is:%

\begin{equation}
C_{2}^{q}=J_{-}J_{+}+\left[  J_{0}\right]  \left[  J_{0}+1\right]
\label{GrindEQ4}%
\end{equation}

with eigenvalues $\left[  I\right]  \left[  I+1\right]  $.

A deformed $q$ like-rotor is a quantum system described by the $su_{q}\left(
2\right)  $ invariant Hamiltonian%

\begin{equation}
H=\frac{\hbar^{2}}{2j^{\left(  0\right)  }}C_{2}^{q}+E_{0},\label{GrindEQ5}%
\end{equation}

where $j^{\left(  0\right)  }$ is the moment of inertia for $q=1$ and $E_{0}$
the bandhead energy. The parameters $j^{\left(  0\right)  }$and $E_{0}$ are
regarded as constants of the model. The rotational energy spectrum can be then
expressed as%

\begin{equation}
E=\frac{\hbar^{2}}{2j^{\left(  0\right)  }}\frac{\sin\left(  I\left\vert
\tau\right\vert \right)  \sin\left[  \left(  I+1\right)  \left\vert
\tau\right\vert \right]  }{\sin^{2}\left\vert \tau\right\vert }+E_{0}%
\label{GrindEQ6}%
\end{equation}

\noindent where here a pure imaginary $\tau\left(  \equiv\ln q=i\left\vert
\tau\right\vert \right)  $is assumed.

An extended version of this model is obtained by replacing $I+1$ by $I+c$
where $c>1$. The addition of the parameter c allows for the description of
nuclear anharmonicities in a way similar to that of the Interacting Boson
Model and the Generalized Variable Moment of Inertia model. The energy
spectrum in this case becomes:%

\begin{equation}
E=\frac{\hbar^{2}}{2j^{\left(  0\right)  }}\frac{\sin\left(  I\left\vert
\tau\right\vert \right)  \sin\left[  \left(  I+c\right)  \left\vert
\tau\right\vert \right]  }{\sin^{2}\left\vert \tau\right\vert }+E_{0}%
,\label{GrindEQ7}%
\end{equation}

\noindent which contains three parameters: the moment of inertia $j^{\left(
0\right)  }$, the deformed parameter $\tau$, and the anaharmonicity parameter
$c$.

In our application of the model in order to fit the three parameters in
equation \eqref{GrindEQ7} we make use of the transition energies of 79 SD
bands in the $A\sim190$ region that are reported for the nuclei Au, Tl, Bi,
Pb, and Po \cite{Singh}. The kinematic $J^{\left(  1\right)  }$ and the
dynamical moment of inertia $J^{\left(  2\right)  }$ are calculated from the
following defining relations:

\noindent\
\begin{equation}
J^{\left(  1\right)  }=[(2I-1)/E_{\gamma}(I)](\hbar^{2}MeV^{-1}%
)\label{GrindEQ8}%
\end{equation}

\begin{equation}
J^{\left(  2\right)  }=4/[E_{\gamma}(I+2)-E_{\gamma}(I)](\hbar^{2}%
MeV^{-1})\label{GrindEQ9}%
\end{equation}

where the transition energy $E_{\gamma}(I)$ is%

\begin{equation}
E_{\gamma}\left(  I\right)  =E\left(  I\right)  -E\left(  I-2\right)
,\label{GrindEQ10}%
\end{equation}

\noindent We have used as a quantitative measure for best fit the root mean
square (rms) $\sigma$ defined as:%

\begin{equation}
\sigma=\sqrt{\frac{1}{N}\sum_{I=1}^{N}\left(  1-\frac{E_{\gamma}^{calc}\left(
I\right)  }{E_{\gamma}^{\exp t}\left(  I\right)  }\right)  ^{2}}%
,\label{GrindEQ11}%
\end{equation}

\noindent where $N$ \ is the number of levels fitted.

\section{RESULTS AND CONCLUSION}

A representative sample of the fitting parameters and the rms of the two
models, for the studied nuclei, are presented in table I and table II. Out of
the 79 studied SD bands 20 of them have the anharmonic parameter $c=1,$ which
we do not include in the tables, since they lead to no comparison between the
two models. These bands are:$^{197}$Bi SD, $^{190}$Hg SD3, $^{191}$Hg SD1,
SD4, $^{192}$Hg SD2, SD3, $^{193}$Hg SD6, $^{195}$Hg SD3, $^{193}$Pb SD2,
$^{195}$Pb SD2, $^{197}$Pb SD2, SD3, SD4, SD5, SD6, $^{192}$Tl SD1, SD2,
$^{191}$Tl SD2, $^{193}$Tl SD3, SD4. In addition 12 SD bands ($^{191}$Au SD3,
$^{189}$Hg SD1, $^{190}$Hg SD4, $^{192}$Hg SD1, $^{194}$Hg SD1, $^{195}$Hg
SD2, $^{190}$Tl SD2, $^{193}$Tl SD5, $^{193}$Pb SD1, $^{195}$Pb SD3, $^{196}%
$Pb SD4, $^{198}$Pb SD2)\ have $c>2$, which are out side the rotational region
\cite{Bonatsos2}. Figs. 1-2 clearly illustrate that our calculations of the
moments of inertia are in good agreement with experimental data at low angular
frequency. Both models give good fit for the kinematic moment of enertia but
they show marked disagreement for the dynamic moment of inertia at high
angular frequency. The models fail to account for the uprising and the down
turn of the dynamic moment of inertia. Comparison of the rms of the studied SD
bands, table I and II, for the two models shows a significant improvement in
favor of the generalized $su_{q}\left(  2\right)  $\noindent.

\bigskip%

\begin{table}%
\caption
{The fitting parameters and rms of the present models (for 1.0 < c < 1.5)}%
\begin{tabular}
[c]{cccccccccc}\hline\hline
$_{{}}$ &  & \multicolumn{3}{c}{su$_{q}\left(  2\right)  $} &  &
\multicolumn{4}{c}{modified su$_{q}\left(  2\right)  $}\\\cline{2-5}%
\cline{7-10}
&  & $\frac{\hbar^{2}}{2j^{\left(  0\right)  }}$ & $\tau$ & $\sigma\%$ &  &
$\frac{\hbar^{2}}{2j^{\left(  0\right)  }}$ & $\tau$ & $c$ & $\sigma
\%$\\\hline
\multicolumn{1}{l}{$^{191}$Au SD1} &  & \multicolumn{1}{l}{5.24206} &
\multicolumn{1}{l}{0.00945} & \multicolumn{1}{l}{0.42239} &
\multicolumn{1}{l}{} & \multicolumn{1}{l}{5.23042} &
\multicolumn{1}{l}{0.00938} & \multicolumn{1}{l}{1.06323} &
\multicolumn{1}{l}{0.41994}\\
\multicolumn{1}{l}{$^{191}$Au SD2} &  & \multicolumn{1}{l}{5.38074} &
\multicolumn{1}{l}{0.01073} & \multicolumn{1}{l}{0.08715} &
\multicolumn{1}{l}{} & \multicolumn{1}{l}{5.32596} &
\multicolumn{1}{l}{0.01041} & \multicolumn{1}{l}{1.38868} &
\multicolumn{1}{l}{0.07117}\\
\multicolumn{1}{l}{$^{190}$Hg SD1} &  & \multicolumn{1}{l}{5.93895} &
\multicolumn{1}{l}{0.01313} & \multicolumn{1}{l}{0.58702} &
\multicolumn{1}{l}{} & \multicolumn{1}{l}{5.84574} &
\multicolumn{1}{l}{0.01275} & \multicolumn{1}{l}{1.45} &
\multicolumn{1}{l}{0.41552}\\
$^{193}$Hg SD1 &  & 5.39158 & 0.01379 & 0.17721 &  & 5.39158 & 0.01276 &
1.02044 & 0.17721\\
$^{193}$Hg SD2 &  & 5.29818 & 0.01004 & 0.52098 &  & 5.22778 & 0.00965 &
1.45 & 0.35778\\
$^{193}$Hg SD4 &  & 5.29818 & 0.01004 & 0.52098 &  & 5.22778 & 0.00965 &
1.45 & 0.35778\\
\multicolumn{1}{l}{$^{193}$Hg SD5} &  & \multicolumn{1}{l}{4.83994} &
\multicolumn{1}{l}{0.00219} & \multicolumn{1}{l}{0.20816} &
\multicolumn{1}{l}{} & \multicolumn{1}{l}{4.81326} & \multicolumn{1}{l}{0.001}
& \multicolumn{1}{l}{1.17459} & \multicolumn{1}{l}{0.19576}\\
\multicolumn{1}{l}{$^{194}$Hg SD2} &  & \multicolumn{1}{l}{5.28511} &
\multicolumn{1}{l}{0.01007} & \multicolumn{1}{l}{0.45569} &
\multicolumn{1}{l}{} & \multicolumn{1}{l}{5.20579} &
\multicolumn{1}{l}{0.00958} & \multicolumn{1}{l}{1.45} &
\multicolumn{1}{l}{0.29270}\\
$^{195}$Hg SD1 &  & \multicolumn{1}{l}{5.26228} & \multicolumn{1}{l}{0.01035}
& \multicolumn{1}{l}{0.76589} & \multicolumn{1}{l}{} &
\multicolumn{1}{l}{5.19947} & \multicolumn{1}{l}{0.01004} &
\multicolumn{1}{l}{1.45 \ \ \ \ } & \multicolumn{1}{l}{0.62843}\\
$^{189}$Tl SD1 &  & \multicolumn{1}{l}{5.52796} & \multicolumn{1}{l}{0.01065}
& \multicolumn{1}{l}{0.04996} & \multicolumn{1}{l}{} &
\multicolumn{1}{l}{5.49779} & \multicolumn{1}{l}{0.01042} &
\multicolumn{1}{l}{1.16635} & \multicolumn{1}{l}{0.03881}\\
$^{189}$Tl SD2 &  & \multicolumn{1}{l}{5.50741} & \multicolumn{1}{l}{0.01098}
& \multicolumn{1}{l}{0.16131} & \multicolumn{1}{l}{} &
\multicolumn{1}{l}{5.45634} & \multicolumn{1}{l}{0.01055} &
\multicolumn{1}{l}{1.26928} & \multicolumn{1}{l}{0.15042}\\
$^{191}$Tl SD1 &  & 5.38536 & 0.01039 & 0.05602 &  & 5.38353 & 0.01037 &
1.01021 & 0.05593\\
$^{192}$Tl SD3 &  & \multicolumn{1}{l}{5.1029} & \multicolumn{1}{l}{0.00890} &
\multicolumn{1}{l}{0.25246} & \multicolumn{1}{l}{} &
\multicolumn{1}{l}{5.03279} & \multicolumn{1}{l}{0.00827} &
\multicolumn{1}{l}{1.40185} & \multicolumn{1}{l}{0.18759}\\
$^{192}$Tl SD4 &  & \multicolumn{1}{l}{5.1066} & \multicolumn{1}{l}{0.00903} &
\multicolumn{1}{l}{0.17178} & \multicolumn{1}{l}{} &
\multicolumn{1}{l}{5.06215} & \multicolumn{1}{l}{0.00861} &
\multicolumn{1}{l}{1.23881} & \multicolumn{1}{l}{0.12532}\\
$^{193}$Tl SD1 &  & \multicolumn{1}{l}{5.18845} & \multicolumn{1}{l}{0.00970}
& \multicolumn{1}{l}{0.27259} & \multicolumn{1}{l}{} &
\multicolumn{1}{l}{5.14984} & \multicolumn{1}{l}{0.00940} &
\multicolumn{1}{l}{1.21010} & \multicolumn{1}{l}{0.24470}\\
\multicolumn{1}{l}{$^{193}$Tl SD2} &  & \multicolumn{1}{l}{5.18638} &
\multicolumn{1}{l}{0.00889} & \multicolumn{1}{l}{0.26542} &
\multicolumn{1}{l}{} & \multicolumn{1}{l}{5.12698} &
\multicolumn{1}{l}{0.00839} & \multicolumn{1}{l}{1.33399} &
\multicolumn{1}{l}{0.21371}\\
$^{194}$Tl SD1 &  & \multicolumn{1}{l}{5.00301} & \multicolumn{1}{l}{0.00835}
& \multicolumn{1}{l}{0.12796} & \multicolumn{1}{l}{} &
\multicolumn{1}{l}{4.97268} & \multicolumn{1}{l}{0.00806} &
\multicolumn{1}{l}{1.18518} & \multicolumn{1}{l}{0.11626}\\
$^{194}$Tl SD2 &  & \multicolumn{1}{l}{5.00398} & \multicolumn{1}{l}{0.00849}
& \multicolumn{1}{l}{0.08714} & \multicolumn{1}{l}{} &
\multicolumn{1}{l}{4.97817} & \multicolumn{1}{l}{0.00821} &
\multicolumn{1}{l}{1.13658} & \multicolumn{1}{l}{0.05851}\\
\multicolumn{1}{l}{$^{195}$Tl SD1} &  & \multicolumn{1}{l}{5.2395} &
\multicolumn{1}{l}{0.00950} & \multicolumn{1}{l}{0.14559} &
\multicolumn{1}{l}{} & \multicolumn{1}{l}{5.23000} &
\multicolumn{1}{l}{0.00942} & \multicolumn{1}{l}{1.04212} &
\multicolumn{1}{l}{0.13952}\\
\multicolumn{1}{l}{$^{195}$Tl SD2} &  & \multicolumn{1}{l}{5.24353} &
\multicolumn{1}{l}{0.01042} & \multicolumn{1}{l}{0.21265} &
\multicolumn{1}{l}{} & \multicolumn{1}{l}{5.21783} &
\multicolumn{1}{l}{0.01023} & \multicolumn{1}{l}{1.12423} &
\multicolumn{1}{l}{0.18672}\\
$^{193}$Pb SD3 &  & \multicolumn{1}{l}{5.2603} & \multicolumn{1}{l}{0.00895} &
\multicolumn{1}{l}{0.19961} & \multicolumn{1}{l}{} &
\multicolumn{1}{l}{5.17564} & \multicolumn{1}{l}{0.00811} &
\multicolumn{1}{l}{1.45755} & \multicolumn{1}{l}{0.11696}\\
$^{193}$Pb SD6 &  & \multicolumn{1}{l}{5.34273} & \multicolumn{1}{l}{0.01056}
& \multicolumn{1}{l}{0.41145} & \multicolumn{1}{l}{} &
\multicolumn{1}{l}{5.25659} & \multicolumn{1}{l}{0.00985} &
\multicolumn{1}{l}{1.45 \ \ \ \ \ } & \multicolumn{1}{l}{0.24225}\\
$^{194}$Pb SD1 &  & \multicolumn{1}{l}{5.62768} & \multicolumn{1}{l}{0.01231}
& \multicolumn{1}{l}{0.72335} & \multicolumn{1}{l}{} &
\multicolumn{1}{l}{5.50272} & \multicolumn{1}{l}{0.01128} &
\multicolumn{1}{l}{1.44987} & \multicolumn{1}{l}{0.40757}\\
$^{194}$Pb SD2 &  & \multicolumn{1}{l}{5.28708} & \multicolumn{1}{l}{0.01133}
& \multicolumn{1}{l}{0.16664} & \multicolumn{1}{l}{} &
\multicolumn{1}{l}{5.23070} & \multicolumn{1}{l}{0.01067} &
\multicolumn{1}{l}{1.25752} & \multicolumn{1}{l}{0.14921}\\
$^{194}$Pb SD3 &  & \multicolumn{1}{l}{5.28808} & \multicolumn{1}{l}{0.01121}
& \multicolumn{1}{l}{0.13644} & \multicolumn{1}{l}{} &
\multicolumn{1}{l}{5.23187} & \multicolumn{1}{l}{0.01061} &
\multicolumn{1}{l}{1.27327} & \multicolumn{1}{l}{0.11896}\\
$^{195}$Pb SD1 &  & \multicolumn{1}{l}{5.05467} & \multicolumn{1}{l}{0.00596}
& \multicolumn{1}{l}{0.16410} & \multicolumn{1}{l}{} &
\multicolumn{1}{l}{5.00077} & \multicolumn{1}{l}{0.00496} &
\multicolumn{1}{l}{1.25433} & \multicolumn{1}{l}{0.06094}\\
$^{195}$Pb SD4 &  & \multicolumn{1}{l}{5.40187} & \multicolumn{1}{l}{0.01143}
& \multicolumn{1}{l}{0.25481} & \multicolumn{1}{l}{} &
\multicolumn{1}{l}{5.37387} & \multicolumn{1}{l}{0.01116} &
\multicolumn{1}{l}{1.12594} & \multicolumn{1}{l}{0.24772}\\
$^{196}$Pb SD1 &  & \multicolumn{1}{l}{5.7067} & \multicolumn{1}{l}{0.01174} &
\multicolumn{1}{l}{0.23161} & \multicolumn{1}{l}{} &
\multicolumn{1}{l}{5.64178} & \multicolumn{1}{l}{0.01124} &
\multicolumn{1}{l}{1.26440} & \multicolumn{1}{l}{0.04316}\\
\multicolumn{1}{l}{$^{196}$Pb SD2} &  & \multicolumn{1}{l}{5.42321} &
\multicolumn{1}{l}{0.01111} & \multicolumn{1}{l}{0.29565} &
\multicolumn{1}{l}{} & \multicolumn{1}{l}{5.34749} &
\multicolumn{1}{l}{0.01039} & \multicolumn{1}{l}{1.34114} &
\multicolumn{1}{l}{0.23212}\\
$^{197}$Pb SD1 &  & 5.09885 & 0.00609 & 0.11375 &  & 5.08275 & 0.00586 &
1.07249 & 0.08923\\
$^{198}$Pb SD1 &  & \multicolumn{1}{l}{5.66046} & \multicolumn{1}{l}{0.00919}
& \multicolumn{1}{l}{0.33644} & \multicolumn{1}{l}{} &
\multicolumn{1}{l}{5.58475} & \multicolumn{1}{l}{0.00871} &
\multicolumn{1}{l}{1.45 \ \ \ \ } & \multicolumn{1}{l}{0.20562}\\\hline\hline
\end{tabular}%
\end{table}%

\bigskip%

\begin{table}%
\caption
{The fitting parameters and rms of the present models (for 1.5 < c < 2.0)}%
\begin{tabular}
[c]{cccccccccc}\hline\hline
$_{{}}$ &  & \multicolumn{3}{c}{su$_{q}\left(  2\right)  $} &  &
\multicolumn{4}{c}{modified su$_{q}\left(  2\right)  $}\\\cline{2-5}%
\cline{7-10}
&  & $\frac{\hbar^{2}}{2j^{\left(  0\right)  }}$ & $\tau$ & $\sigma\%$ &  &
$\frac{\hbar^{2}}{2j^{\left(  0\right)  }}$ & $\tau$ & $c$ & $\sigma\%$\\
\multicolumn{1}{l}{$^{198}$Po SD} &  & \multicolumn{1}{l}{5.87348} &
\multicolumn{1}{l}{0.015368} & \multicolumn{1}{l}{0.37078} &
\multicolumn{1}{l}{} & \multicolumn{1}{l}{5.71292} &
\multicolumn{1}{l}{0.01359} & \multicolumn{1}{l}{1.52075} &
\multicolumn{1}{l}{0.10810}\\
\multicolumn{1}{l}{$^{196}$Bi SD} &  & \multicolumn{1}{l}{5.47304} &
\multicolumn{1}{l}{0.009596} & \multicolumn{1}{l}{0.61989} &
\multicolumn{1}{l}{} & \multicolumn{1}{l}{5.26217} &
\multicolumn{1}{l}{0.00646} & \multicolumn{1}{l}{1.82055} &
\multicolumn{1}{l}{0.03724}\\
$^{191}$Hg SD2 &  & \multicolumn{1}{l}{5.28001} & \multicolumn{1}{l}{0.009649}
& \multicolumn{1}{l}{0.23706} & \multicolumn{1}{l}{} &
\multicolumn{1}{l}{5.18865} & \multicolumn{1}{l}{0.00899} &
\multicolumn{1}{l}{1.54006} & \multicolumn{1}{l}{0.08739}\\
$^{191}$Hg SD3 &  & \multicolumn{1}{l}{5.27424} & \multicolumn{1}{l}{0.010148}
& \multicolumn{1}{l}{0.31120} & \multicolumn{1}{l}{} &
\multicolumn{1}{l}{5.15491} & \multicolumn{1}{l}{0.00937} &
\multicolumn{1}{l}{1.75194} & \multicolumn{1}{l}{0.15758}\\
$^{193}$Hg SD3 &  & \multicolumn{1}{l}{5.308} & \multicolumn{1}{l}{0.010110} &
\multicolumn{1}{l}{0.54763} & \multicolumn{1}{l}{} &
\multicolumn{1}{l}{5.15232} & \multicolumn{1}{l}{0.00927} &
\multicolumn{1}{l}{1.95203} & \multicolumn{1}{l}{0.25438}\\
\multicolumn{1}{l}{$^{194}$Hg SD3} &  & \multicolumn{1}{l}{5.26475} &
\multicolumn{1}{l}{0.009839} & \multicolumn{1}{l}{0.55888} &
\multicolumn{1}{l}{} & \multicolumn{1}{l}{5.11966} &
\multicolumn{1}{l}{0.00897} & \multicolumn{1}{l}{1.88759} &
\multicolumn{1}{l}{0.25922}\\
$^{195}$Hg SD4 &  & \multicolumn{1}{l}{5.08002} & \multicolumn{1}{l}{0.008086}
& \multicolumn{1}{l}{0.21760} & \multicolumn{1}{l}{} &
\multicolumn{1}{l}{4.98767} & \multicolumn{1}{l}{0.00753} &
\multicolumn{1}{l}{1.73243} & \multicolumn{1}{l}{0.13646}\\
\multicolumn{1}{l}{$^{192}$Pb SD} &  & \multicolumn{1}{l}{5.73047} &
\multicolumn{1}{l}{0.013809} & \multicolumn{1}{l}{0.56276} &
\multicolumn{1}{l}{} & \multicolumn{1}{l}{5.5929} &
\multicolumn{1}{l}{0.01268} & \multicolumn{1}{l}{1.57881} &
\multicolumn{1}{l}{0.47779}\\
$^{193}$Pb SD4 &  & \multicolumn{1}{l}{5.30025} & \multicolumn{1}{l}{0.010653}
& \multicolumn{1}{l}{0.22269} & \multicolumn{1}{l}{} &
\multicolumn{1}{l}{5.18738} & \multicolumn{1}{l}{0.00978} &
\multicolumn{1}{l}{1.65015} & \multicolumn{1}{l}{0.09640}\\
$^{193}$Pb SD5 &  & \multicolumn{1}{l}{5.34677} & \multicolumn{1}{l}{0.010707}
& \multicolumn{1}{l}{0.36132} & \multicolumn{1}{l}{} &
\multicolumn{1}{l}{5.22763} & \multicolumn{1}{l}{0.00965} &
\multicolumn{1}{l}{1.58574} & \multicolumn{1}{l}{0.20445}\\
$^{196}$Pb SD3 &  & \multicolumn{1}{l}{5.4111} & \multicolumn{1}{l}{0.011044}
& \multicolumn{1}{l}{0.29265} & \multicolumn{1}{l}{} &
\multicolumn{1}{l}{5.29645} & \multicolumn{1}{l}{0.01002} &
\multicolumn{1}{l}{1.55773} & \multicolumn{1}{l}{0.15156}\\
$^{198}$Pb SD3 &  & \multicolumn{1}{l}{5.68231} & \multicolumn{1}{l}{0.011113}
& \multicolumn{1}{l}{0.32830} & \multicolumn{1}{l}{} &
\multicolumn{1}{l}{5.54283} & \multicolumn{1}{l}{0.00982} &
\multicolumn{1}{l}{1.60574} & \multicolumn{1}{l}{0.07149}\\
$^{194}$Tl SD3 &  & \multicolumn{1}{l}{5.22953} & \multicolumn{1}{l}{0.009804}
& \multicolumn{1}{l}{0.30147} & \multicolumn{1}{l}{} &
\multicolumn{1}{l}{5.09825} & \multicolumn{1}{l}{0.00864} &
\multicolumn{1}{l}{1.72394} & \multicolumn{1}{l}{0.08790}\\
$^{194}$Tl SD4 &  & \multicolumn{1}{l}{5.23294} & \multicolumn{1}{l}{0.009887}
& \multicolumn{1}{l}{0.37686} & \multicolumn{1}{l}{} &
\multicolumn{1}{l}{5.08547} & \multicolumn{1}{l}{0.00847} &
\multicolumn{1}{l}{1.76634} & \multicolumn{1}{l}{0.12778}\\
$^{194}$Tl SD5 &  & \multicolumn{1}{l}{4.93312} & \multicolumn{1}{l}{0.008534}
& \multicolumn{1}{l}{0.31105} & \multicolumn{1}{l}{} &
\multicolumn{1}{l}{4.81538} & \multicolumn{1}{l}{0.00685} &
\multicolumn{1}{l}{1.58204} & \multicolumn{1}{l}{0.06564}\\
$^{194}$Tl SD6 &  & \multicolumn{1}{l}{4.93169} & \multicolumn{1}{l}{0.008057}
& \multicolumn{1}{l}{0.24610} & \multicolumn{1}{l}{} &
\multicolumn{1}{l}{4.83214} & \multicolumn{1}{l}{0.00656} &
\multicolumn{1}{l}{1.50663} & \multicolumn{1}{l}{0.12430}\\\hline\hline
\end{tabular}%
\end{table}%

\bigskip\bigskip%
\begin{figure}
[ptb]
\begin{center}
\includegraphics[
height=3.8605in,
width=3.2811in
]%
{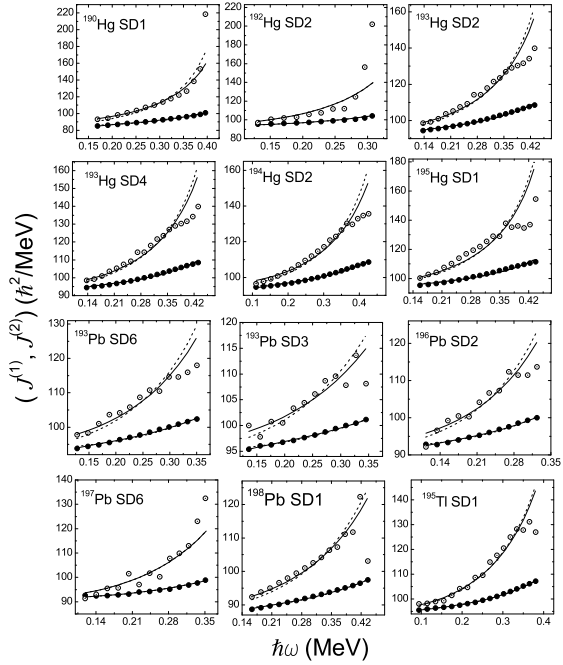}%
\caption{Comparison between the experimental and theoretical calculation of
the Kinematic \textit{J}$^{\left(  1\right)  }$ (dote) and Dynamic
\textit{J}$^{\left(  2\right)  }$(circle) moments of inertia versus the
rotational frequency ( $\hbar\omega$) of a representetive sample of
superdeformed bands in the $A\sim190$ region. The modified su$_{q}$(2) model
(full line) and the nonmodified su$_{q}$(2) (dashed line). $^{192}$Hg SD2 and
$^{197}$Pb SD6 have $c=1$, and $c$ for the rest SDs as in table I.}%
\end{center}
\end{figure}
%

\begin{figure}
[ptb]
\begin{center}
\includegraphics[
height=3.8605in,
width=3.1834in
]%
{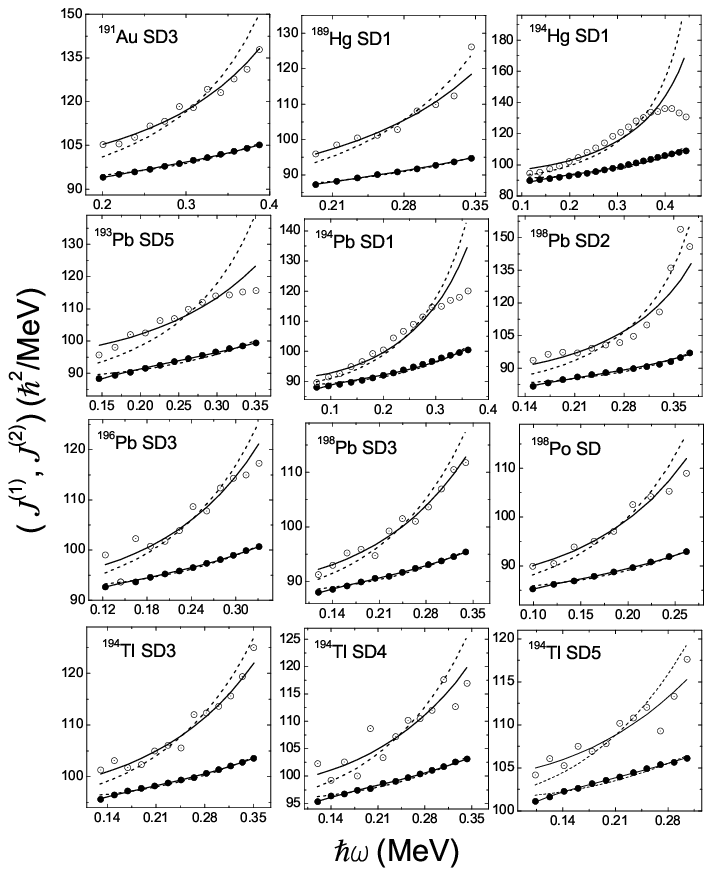}%
\caption{Comparison between the experimental and theoretical calculation of
the Kinematic \textit{J}$^{\left(  1\right)  }$ (dote) and Dynamic
\textit{J}$^{\left(  2\right)  }$(circle) moments of inertia versus the
rotational frequency ( $\hbar\omega$) of a representetive sample of
superdeformed bands in the $A\sim190 $ region. The modified su$_{q}$(2) model
(full line) and the nonmodified su$_{q}$(2) (dashed line). $^{191}$Au SD3,
$^{189}$Hg SD1, $^{194}$Hg SD1, $^{198}$Pb SD2 have
$c=3.35953,2.39011,2.47371,2.81838$ respectively; $^{194}$Pb SD1 has $c$ as in
table I; and $c$ for the rest as in table II.}%
\end{center}
\end{figure}

\begin{acknowledgement}
We would like to thank the referee for his valuable remarks.
\end{acknowledgement}

\end{document}